\documentclass{article}
\usepackage{spconf,amsmath,graphicx,amssymb,graphicx}
\usepackage{booktabs,pifont,cite}
\usepackage{caption}
\usepackage{multirow}

\title{Automatic channel selection and spatial feature integration for multi-channel speech recognition across various array topologies}
\name{
\begin{tabular}{c}
  \it Bingshen Mu$^1$, Pengcheng Guo$^1$, Dake Guo$^1$, Pan Zhou$^2$, Wei Chen$^2$, Lei Xie$^{1*}$\thanks{$^*$: Corresponding author.}
\end{tabular}
}

\address{
  $^1$Audio, Speech and Language Processing Group (ASLP@NPU), School of Computer Science, \\Northwestern Polytechnical University, Xian, China\\
  $^2$Space AI, Li Auto
}
\begin{document}
\ninept
\maketitle
\begin{abstract}
\vspace{-0.1cm}
% Automatic Speech Recognition (ASR) has shown remarkable progress, yet challenges persist in real-world scenarios marked by complex microphone arrays. 
Automatic Speech Recognition (ASR) has shown remarkable progress, yet it still faces challenges in real-world distant scenarios across various array topologies each with multiple recording devices.
The focal point of the CHiME-7 Distant ASR task is to devise a unified system capable of generalizing various array topologies that have multiple recording devices and offering reliable recognition performance in real-world environments. 
Addressing this task, we introduce an ASR system that demonstrates exceptional performance across various array topologies.
First of all, we propose two attention-based automatic channel selection modules to select the most advantageous subset of multi-channel signals from multiple recording devices for each utterance.
Furthermore, we introduce inter-channel spatial features to augment the effectiveness of multi-frame cross-channel attention, aiding it in improving the capability of spatial information awareness.
Finally, we propose a multi-layer convolution fusion module drawing inspiration from the U-Net architecture to integrate the multi-channel output into a single-channel output.
Experimental results on the CHiME-7 corpus with oracle segmentation demonstrate that the improvements introduced in our proposed ASR system lead to a relative reduction of 40.1\% in the Macro Diarization Attributed Word Error Rates (DA-WER) when compared to the baseline ASR system on the Eval sets.
% Addressing this task, we introduce an ASR system that excels consistently across diverse array topologies. 
% Our ASR system seamlessly incorporates two attention-based automatic channel selection modules into the existing ASR architecture. These modules automatically select the most advantageous subset of channels from multiple signal channels, thereby mitigating the detrimental impact of noise-contaminated channels on ASR. 
% Automatic channel selection is used to select the most advantageous subset of multi-channel signals from multiple recording devices for each utterance, thereby mitigating the negative impact of noisy channels on ASR. 
% To select the most advantageous subset of multi-channel signals from multiple recording devices for each utterance, we propose two attention-based automatic channel selection modules that can be seamlessly integrated into the existing ASR system.
% Furthermore, we introduce inter-channel spatial features to augment the effectiveness of multi-frame cross-channel attention, aiding the model in improving its spatial information perception capability.
% Finally, we propose an architecture employing a multi-layered convolutional module inspired by the U-Net design, to integrate the outputs from multiple channels.
% Pertinent to mention is the fact that all the improvements we propose are directed solely towards the ASR system, no speech enhancement model is used. 

\end{abstract}
\vspace{-0.1cm}
\section{Introduction} \label{sec:intro}
\vspace{-0.1cm}
% With the advancements in deep learning, automatic speech recognition (ASR) has made significant progress, leading to noticeable improvements across various speech applications.
As deep learning continues to advance, automatic speech recognition (ASR) technology has made significant progress, leading to substantial improvements in its performance.
However, ASR systems continue to encounter challenges in real-world distant scenarios characterized by factors like background noise, reverberation, speaker overlap, and various array topologies.
To tackle these challenges, the CHiME Challenge series~\cite{barker2013pascal, vincent2013second, barker2015third, vincent20164th, barker18_interspeech, watanabe2020chime, cornell2023chime} has been established to boost the development of robust ASR systems by promoting research and innovation in multi-microphone signal processing algorithms.

The CHiME-7 Distant ASR (DASR) task this year focuses on designing a system that can generalize across various array geometries (e.g., linear, circle, and ad-hoc array) and provide reliable recognition performance in a wide range of real-world settings, even under adverse acoustic conditions~\cite{cornell2023chime}.
In this task, multiple recording devices are used to capture audio from different spatial locations simultaneously, enabling a better coverage of the sound source.
% However, taking advantage of the information from different channels effectively remains a challenge.
However, when the number of recording devices becomes large, it may not be advisable to incorporate all devices into the analysis given that the audio recorded by some devices is affected heavily by background noise, resulting in significant degradation of ASR performance.
Besides, taking advantage of the information from devices located in different spatial locations effectively remains a challenge.
Furthermore, it is worthwhile to explore the adaptation of an ASR system to various array topologies without using prior array-specific information.
% Besides, it is worthwhile to explore the adaptation of an ASR system to various array geometries without using prior array-specific information.
% Besides, some of the far-field microphone arrays or channels may be contaminated by background noise, resulting in significant degradation of ASR performance.
% Furthermore, certain distant microphone arrays or channels are affected by background noise, resulting in significant degradation of ASR performance.

% 之前的通道选择方法介绍
Automatic channel selection is an effective approach to selecting the most promising subset of multi-channel signals for each utterance. One of the advantages of channel selection is that it can generalize across different array topologies. Previous channel selection measures may be classified into two groups: signal-based and decoder-based measures. Signal-based measures include position and orientation methods~\cite{wolf2010potential}, energy and signal-to-noise ratio methods~\cite{obuchi2004multiple, wolf2010potential}, room impulse response methods~\cite{wolf2009towards, jeub2011blind}, and envelope-variance methods~\cite{wolf2010potential, wolf2014channel}. Decoder-based measures include likelihood~\cite{shimizu2000speech}, pairwise likelihood normalization~\cite{wolf2012pairwise}, feature normalization~\cite{obuchi2004multiple}, and class separability~\cite{wolfel2007channel}.
% 引入我们的通道选择方法
Conventional channel selection methods require preprocessing the audio or post-processing the ASR results, consequently resulting in protracted and inflexible processing pipelines. Hence, we propose two attention-based automatic channel selection modules: coarse-grained channel selection (CGCS) and fine-grained channel selection (FGCS). 
These two modules can be integrated into the ASR system and learn to assign higher weights to channels or frames that are beneficial for ASR performance. 
% while assigning lower weights to channels or frames that are detrimental to ASR. 
Due to the remarkable performance of the guided source separation (GSS)~\cite{boeddeker2018front} algorithm, we utilize the single-channel audio processed by the GSS algorithm as the target for channel selection. Additionally, we replace the conventional residual connection between the original multi-channel audio features and the channel-selected multi-channel audio features with the gated residual connection (GRC). This change prevents the conventional residual connection from reintroducing original multi-channel audio features that are not conducive to ASR performance back into the channel-selected multi-channel audio features and allows autonomous learning of which features from the original multi-channel audio features should be incorporated into the channel-selected multi-channel audio features.
% Instead, it allows the GRC to autonomously learn which features from the original multi-channel audio features should be incorporated into the channel-selected multi-channel audio features.

% MFCCA介绍
% The attention mechanism recently has been incorporated into neural beamforming~\cite{chang2020end, tolooshams2020channel} to perform recursive non-linear beamforming on data in latent spaces. 
% Conspicuously, cross-channel attention has been introduced to directly harness the potential inherent in multi-channel signals within ASR systems~\cite{chang2021end, chang2021multi}. 
Cross-channel attention has recently been introduced in ASR systems to directly harness the inherent potential of multi-channel signals~\cite{chang2021end, chang2021multi}.
% Cross-channel attention recently has been introduced to directly harness the potential inherent in multi-channel signals within ASR systems~\cite{chang2021end, chang2021multi}. 
It is truly remarkable that this approach can bypass the complicated front-end formalization and seamlessly incorporate beamforming and acoustic modeling into an end-to-end neural network. Furthermore, the cross-channel attention is agnostic to the number and topology configuration of arrays, making it particularly well-suited for the demands of the CHiME-7 task. This cross-channel attention approach takes frame-level multi-channel signals as input and learns the global correlations between different channels. 
Multi-Frame Cross-Channel Attention (MFCCA)~\cite{yu2023mfcca} models cross-channel information between adjacent frames while leveraging both channel and frame information. 
% It can be straightforwardly described as mapping each channel representation (query) to a set of channel-averaged representations (key-value) to produce an output~\cite{chang2021end, chang2021multi}, known as Frame-Level Cross-Channel Attention (FLCCA). Additionally, Channel-Level Cross-Channel Attention (CLCCA) is computed along the channel dimension, combining the representation of each channel with those of other channels within each time step~\cite{wang2020neural}, functioning similarly to beamforming. 
% Multi-Frame Cross-Channel Attention (MFCCA)~\cite{yu2023mfcca} capitalizes on the complementarity between frame-level and channel-level cross-channel attention, modeling cross-channel information between adjacent frames while leveraging both channel and frame information. 
% 引入我们的MFCCA改进方法
Given that MFCCA implicitly models spatial information between different channels through the attention mechanism, we propose an improvement to MFCCA by incorporating additional inter-channel spatial features, such as inter-channel phase difference (IPD)~\cite{yoshioka2018multi, gu2019end}. These spatial features explicitly guide the MFCCA in modeling spatial information and capturing desired signals while suppressing interfering sources. 

% 多通道融合模块的介绍和改进
To integrate multi-channel outputs, previous research~\cite{chang2021multi, wang2022cross} often averages or concatenates features along the channel dimension. However, directly reducing the channel dimension can lead to the loss of channel-specific information in multi-channel outputs. 
To address this, we employ a multi-layer convolution fusion module based on U-Net~\cite{ronneberger2015u} to gradually reduce the channel dimension,  transforming the multi-channel output into a single-channel output. 
% a multi-layer convolution module drawing inspiration from the U-Net architecture
% To address this, we employ a U-Net~\cite{ronneberger2015u} architecture with multiple convolution layers that gradually reduce the channel dimension, ultimately transforming the multi-channel output into a single-channel output. 
Our convolution fusion module employs skip connections and the fusion of multi-scale features, mitigating the loss of channel-specific information as the channel dimension gradually decreases.

% intro 总结部分
In summary, we propose two attention-based automatic channel selection modules with GRC to select the most promising subset of multi-channel signals for ASR. Furthermore, we propose an improvement to MFCCA by incorporating inter-channel spatial features, enabling MFCCA to perceive spatial relationships between different channels more clearly. Additionally, we improve the multi-channel output fusion module by employing a U-Net-based convolution fusion module to more effectively integrate the multi-channel output. 
% It is imperative to highlight that all the improvements we propose are directed solely towards the ASR system. 
Experiments conducted on the CHiME-7 corpus with oracle segmentation indicate that the improvements made to our proposed ASR system result in a relative reduction of 40.1\% in Macro Diarization Attributed Word Error Rates (DA-WER)~\cite{cornell2023chime} compared to the baseline ASR system on the Eval sets.

\vspace{-9pt}
\section{Proposed system}
\vspace{-9pt}
\subsection{Data processing} \label{sec:data_process}
\begin{figure}[!t]
  \centering
  \includegraphics[width=7.5cm]{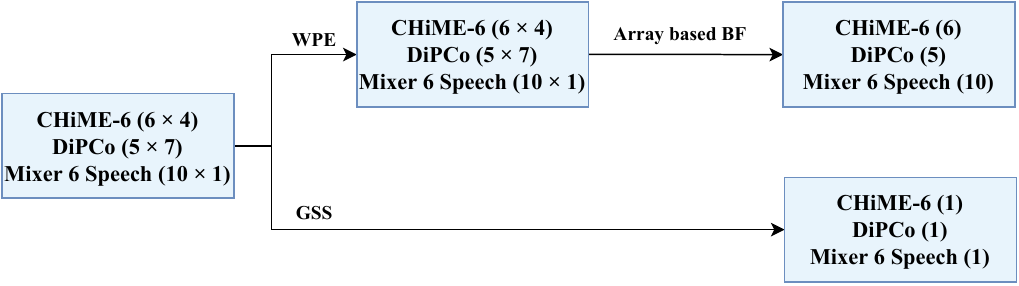}
  \caption{The flow chart of data processing. ($N$ × $M$) denotes $N$ array devices each with $M$ microphones. ($K$) denotes $K$ channels after data processing.}
  \vspace{-18pt}
   \label{fig:data_processing}
\end{figure}

\vspace{-0.1cm}
 % CHiME-7 DASR is composed of three different datasets: CHiME-6~\cite{watanabe2020chime} whose scenario is the dinner party and recordings from binaural microphones worn by each speaker are provided along with distant speech captured by 6 Kinect array devices with 4 microphones each for a total of 24 microphones, DiPCo~\cite{van2019dipco} whose scenario is the dinner party and recordings from close microphones are provided along with distant speech captured by 5 far-field devices with a 7-mic circular array each for a total of 35 microphones, and Mixer 6 Speech~\cite{brandschain2010mixer} whose scenario is the meeting scenario and recordings are all captured by 14 microphones of varying styles. 
CHiME-7 DASR task is composed of three different datasets: CHiME-6~\cite{watanabe2020chime} whose distant speech captured by 6 Kinect array devices with 4 microphones each for a total of 24 microphones, DiPCo~\cite{van2019dipco} whose distant speech captured by 5 far-field devices with a 7-mic circular array each for a total of 35 microphones, and Mixer 6 Speech~\cite{brandschain2010mixer} whose distant speech are captured by 10 microphones of varying styles.
Given the substantial microphone counts present in each dataset, preprocessing of the data becomes imperative. Fig.~\ref{fig:data_processing} shows our data processing progress. 
These three datasets are first preprocessed using the weighted prediction error (WPE)~\cite{drude2018nara} and GSS algorithms to obtain enhanced signals for each utterance. After WPE processing, the multi-channel audio from each array is transformed into single-channel audio using the array-based BeamformIt~\cite{anguera2007acoustic} algorithm. 
As a result, the multi-channel audio from multiple array devices is converted into one multi-channel audio, where the number of channels equals the number of arrays.
% As a result, the multi-array multi-channel audio is converted into single multi-channel audio,
% where the number of channels remains the same as the number of arrays.
% where the number of channels equals the number of arrays.
% This data processing approach is applied to both the train, dev, and eval sets.

% \vspace{-0.1cm}

\vspace{-9pt}
\subsection{Attention-based CGCS with GRC}
\begin{figure*}[!t]
  \centering
  \includegraphics[width=0.9\linewidth]{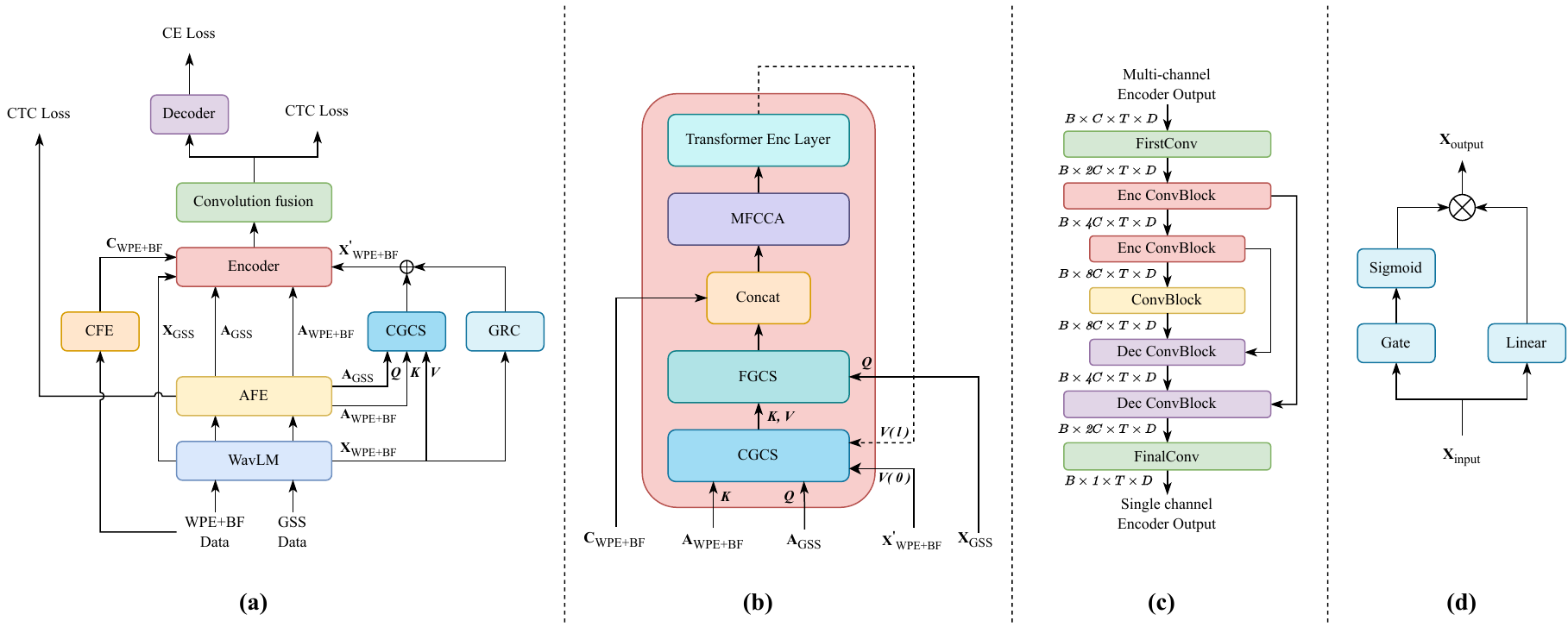}
  \caption{(a) An overview of our proposed ASR system.
  % , where ``AFE" denotes the audio feature extractor and ``CFE" denotes the cosIPD feature extractor. 
  (b) A detailed description of each Encoder layer, where the subscript `` \emph{l} " denotes layer index. 
  (c) The architecture of Convolution fusion. (d) The architecture of GRC, `` Gate '' is a Linear layer.}
  \vspace{-9pt}
  \label{fig:model_structure}
\end{figure*}

\vspace{-0.1cm}
% The attention mechanism assigns higher weights to channels that have abundant semantic information, while assigning lower weights to channels that lack semantic information.
We think that the richness of audio semantic information is correlated with the noise it contains. The more noise an audio contains, the more interference it poses to the human voice, consequently the less semantic information is retained. The CGCS is performed based on the richness of semantic information contained in each channel of the multi-channel audio. We utilize the Gated Recurrent Unit (GRU)~\cite{cho2014learning} network as the channel-level audio feature extractor (AFE), taking the final hidden state as the feature representation for each channel. To extract semantically relevant channel-level audio features, we employ the CTC loss function to guide the AFE. Fig.~\ref{fig:model_structure} illustrates the input for CGCS. The query $\mathbf{A}_{\text{GSS}}\in \mathbb{R}^{B\times 1\times 1\times D}$ and key $\mathbf{A}_{\text{WPE+BF}}\in \mathbb{R}^{B\times K\times 1\times D}$ in the attention mechanism are features extracted from the GSS and WPE+BF audio by the WavLM~\cite{chen2022wavlm} and AFE, respectively.
For the CGCS outside the encoder, the value is $\mathbf{X}_{\text{WPE+BF}}\in \mathbb{R}^{B\times K\times T\times D}$ extracted from the WPE+BF audio by the WavLM. Subsequently, the output of this CGCS will be added with $\mathbf{X}_{\text{WPE+BF}}$ through GRC, resulting in the preliminary selected feature $\mathbf{X^{\prime}}_{\text{WPE+BF}}\in \mathbb{R}^{B\times K\times T\times D}$.
GRC here is to autonomously learn which features from $\mathbf{X}_{\text{WPE+BF}}$ should be incorporated into $\mathbf{X^{\prime}}_{\text{WPE+BF}}$.
For the CGCS in each encoder layer, the value $V(l)$ is the output of the previous encoder layer, while the value $V(0)$ is the preliminary selected feature $\mathbf{X^{\prime}}_{\text{WPE+BF}}$ when $l$ equals 0. In CGCS, the query should be repeated along the time and channel dimensions denoted as $\mathbf{A^{\prime}}_{\text{GSS}}\in \mathbb{R}^{B\times K\times T\times D}$, and the key should be repeated along the time dimension denoted as $\mathbf{A^{\prime}}_{\text{WPE+BF}}\in \mathbb{R}^{B\times K\times T\times D}$. The output of CGCS in the first encoder layer is calculated as
\begin{equation}
    \begin{aligned}
        &\mathbf{Q}^{CGCS} = \mathbf{A^{\prime}}_{\text{GSS}}\mathbf{W}^{CGCS,q} + \left( \mathbf{b}^{CGCS,q} \right ) ^{T},\\
        &\mathbf{K}^{CGCS} = \mathbf{A^{\prime}}_{\text{WPE+BF}}\mathbf{W}^{CGCS,k}  + \left( \mathbf{b}^{CGCS,k} \right ) ^{T},\\
        &\mathbf{V}^{CGCS} = \mathbf{X^{\prime}}_{\text{WPE+BF}}\mathbf{W}^{CGCS,v} + \left( \mathbf{b}^{CGCS,v} \right ) ^{T}, \\
        &\mathbf{H}^{CGCS} = \textit{softmax} \left (\frac{\left (\mathbf{Q}^{CGCS}\mathbf{K}^{CGCS}\right )^{T}}{\sqrt{D}}  \right )\mathbf{V}^{CGCS},
    \end{aligned}
\end{equation}
where $\mathbf{W}^{CGCS,*}$ and $\mathbf{b}^{CGCS,*}$ are learnable weight and bias parameters respectively.

% \vspace{-0.1cm}

% \vspace{-0.1cm}
\vspace{-9pt}
\subsection{Attention-based FGCS}
\vspace{-0.1cm}
% CGCS leverages the semantic information of the entire audio from each channel to filter the multi-channel audio features, selecting channels that encompass abundant semantic information. 
CGCS leverages the channel-level semantic information to select channels that encompass abundant semantic information. 
In contrast, FGCS focuses on calculating the frame-level similarity between the GSS audio features and the multi-channel audio features. It assigns higher weights to the frames of multi-channel audio features that are similar to each frame of the GSS audio features. Fig.~\ref{fig:model_structure} illustrates the input for FGCS. The query $\mathbf{X}_{\text{GSS}}\in \mathbb{R}^{B\times 1\times T\times D}$ in the attention mechanism is the feature extracted from the GSS audio by the WavLM, while the key and value are obtained from the output of CGCS. In FGCS, the query should be repeated along the channel dimension denoted as $\mathbf{X^{\prime}}_{\text{GSS}}\in \mathbb{R}^{B\times K\times T\times D}$. The output of FGCS is calculated as
\begin{equation}
    \begin{aligned}
        &\mathbf{Q}^{FGCS} = \mathbf{X^{\prime}}_{\text{GSS}}\mathbf{W}^{FGCS,q} + \left( \mathbf{b}^{FGCS,q} \right ) ^{T},\\
        &\mathbf{K}^{FGCS} = \mathbf{H}^{CGCS}\mathbf{W}^{FGCS,k} + \left( \mathbf{b}^{FGCS,k} \right ) ^{T},\\
        &\mathbf{V}^{FGCS} = \mathbf{H}^{CGCS}\mathbf{W}^{FGCS,v}  + \left( \mathbf{b}^{FGCS,v} \right ) ^{T}, \\
        &\mathbf{H}^{FGCS} = \textit{softmax} \left (\frac{\left (\mathbf{Q}^{FGCS}\mathbf{K}^{FGCS}\right )^{T}}{\sqrt{D}}  \right )\mathbf{V}^{FGCS},
    \end{aligned}
\end{equation}
where $\mathbf{W}^{FGCS,*}$ and $\mathbf{b}^{FGCS,*}$ are learnable weight and bias parameters respectively.
\vspace{-9pt}
\subsection{MFCCA with inter-channel spatial features}
\vspace{-0.1cm}
The spatial information is of vital importance for the multi-channel scenario. Nevertheless, MFCCA lacks explicit spatial features as inputs, relying solely on multi-channel audio features for input, thereby enabling MFCCA to implicitly learn spatial information. 
With the aim of improving the capability of spatial information awareness, we incorporate the inter-channel spatial features named cosIPD features into MFCCA. Specifically, we concatenate the cosIPD features with the output of FGCS and utilize the MFCCA to better perceive spatial information. Specifically, we concatenate $\mathbf{H}^{FGCS}$ with the cosIPD features that are repeated along the time dimension $\mathbf{C}_{\text{WPE+BF}}\in \mathbb{R}^{B\times K\times T\times D}$, extracted by the cosIPD feature extractor (CFE) based on GRU from the WPE+BF audio.
\vspace{-9pt}
\subsection{Convolution fusion}
\vspace{-0.1cm}
To integrate the multi-channel outputs, prior research~\cite{chang2021multi, wang2022cross} mainly averages or merges features along the channel dimension.
In order to alleviate the adverse impact of directly reducing channel dimension and thus preserving channel-specific information,~\cite{yu2023mfcca} employs a straightforward stack of multiple convolution layers to gradually decrease the channel dimension. 
However, it can lead to channel-specific information loss as the channel dimension is progressively reduced, particularly after multiple convolution layers of stacking. 
We improve the convolution fusion module inspired by the architecture of U-Net. Fig.~\ref{fig:model_structure}(c) illustrates the structure of our convolution fusion module.
The U-Net-based convolution fusion module employs skip connections and the fusion of multi-scale features.
Skip connections allow for the direct transfer of channel-specific information between the U-Net encoders and decoders, aiding in preventing channel-specific information loss during the propagation process in stacked multiple convolution layers.
Moreover, the module combines multi-channel audio features from different channel dimensions, enabling it to capture channel-specific information at multiple scales.
% enabling the exchange of information between low-level features and high-level features across the U-Net encoder and decoder. This facilitates the preservation of finer details in the decoder while providing a richer contextual understanding.
% Skip connections help alleviate this issue of information loss by reintroducing earlier encoder features into the decoder.
Each ConvBlock consists of a 2-D convolution layer, Layer Normalization, and PReLU activation function.
The input channel number $C$ in the convolution fusion module remains fixed (i.e., $C$ equals 10 in our work). Consequently, if the channel number of multi-channel encoder output $K$ is smaller than the preconfigured value $C$, expansion of channels is achieved by simple repeating. 
% \subsection{Inference procedure}
% % \vspace{-0.1cm}
% During the inference, the dev and eval sets are first segmented by the baseline SD model results for the main track and oracle diarization for the sub-track, enhanced by WPE, BF, and GSS, transcribed by our ASR models, and rescored by a Transformer based language model (LM) trained on a combination of the CHiME-7 and LibriSpeech~\cite{panayotov2015librispeech} corpora. We also tune the decoding parameters, including beam size, CTC weight, and LM weight, during the decoding process. The results from different models are fused by the ROVER technique finally. 
\vspace{-9pt}
\section{Experiments}
\vspace{-0.1cm}
\begin{table*}[t]
    \caption{The DA-WER(\%) results of our proposed ASR systems on the Dev and Eval sets.}
    \label{tab:all_results}
    \centering
\begin{tabular}{lcccccccc}
\toprule
\multicolumn{1}{c}{\multirow{2}{*}{\textbf{ASR system}}} & \multicolumn{3}{c}{\textbf{Dev Scenario}}     & \multirow{2}{*}{\textbf{Dev Macro}} & \multicolumn{3}{c}{\textbf{Eval Scenario}}    & \multirow{2}{*}{\textbf{Eval Macro}} \\ \cmidrule{2-4} \cmidrule{6-8}
\multicolumn{1}{c}{}                                     & \multicolumn{1}{l}{CHiME-6}  & \multicolumn{1}{l}{DiPCo}   & \multicolumn{1}{l}{Mixer 6}       &    & \multicolumn{1}{l}{CHiME-6}  & \multicolumn{1}{l}{DiPCo}      & \multicolumn{1}{l}{Mixer 6}       &                                      \\ \midrule
Baseline                                                 & 32.6          & 33.5          & 20.2          & 28.8                                & 35.5          & 36.3          & 28.6          & 33.4                                 \\
MFCCA                                                    & 31.1          & 34.3          & 21.7          & 29.0                                & 36.8          & 37.5          & 17.1          & 30.4                                 \\
\hspace{1em}+CGCS                                                    & 28.8          & 30.5          & 18.6          & 26.0                                & 33.3          & 31.7          & 15.3          & 26.8                                 \\
\hspace{2em}+GRC                                                     & 26.5          & 30.8          & 17.7          & 25.0                                & 27.2          & 31.7          & 18.2          & 25.7                                 \\
\hspace{1em}+FGCS                                                    & 28.2          & 30.4          & 16.1          & 24.9                                & 32.7          & 28.6          & 15.4          & 25.6                                 \\
\hspace{1em}+cosIPD                                                  & 29.1          & 30.7          & 15.9          & 25.2                                & 33.4          & 31.2          & 15.4          & 26.4                                 \\
\hspace{1em}+U-NetFusion                                              & 28.7          & 30.4          & 18.5          & 25.9                                & 33.2          & 31.6          & 15.3          & 26.7                                 \\
\hspace{1em}+ALL                                                     & 24.5          & 26.4          & 14.0          & 21.7                                & 28.1          & 24.6          & 13.3          & 22.0                                 \\ \midrule
ROVER                                                    & \textbf{22.8} & \textbf{24.5} & \textbf{13.0} & \textbf{20.1}                       & \textbf{25.6} & \textbf{22.3} & \textbf{12.0} & \textbf{20.0}                        \\ \bottomrule
\vspace{-18pt}
\end{tabular}
\end{table*}
\vspace{-9pt}
\subsection{Baseline}
\vspace{-0.1cm}
Our baseline is identical to the baseline of the CHiME-7 DASR task~\cite{cornell2023chime}. It is noteworthy that the diarizer component is omitted since oracle segmentation is provided. The baseline uses automatic channel selection with an envelope-variance method~\cite{wolf2014channel} for later processing via GSS.
For the baseline ASR, we directly take the model from~\cite{masuyama2023end, chang2022end}, which consists of a hybrid CTC/Attention Transformer~\cite{vaswani2017attention} encoder-decoder ASR model with pretrained and frozen WavLM feature extractor.
% It is trained on the full CHiME-7 train sets. In addition, we used the augmentation scheme as described in the CHiME-6 baseline~\cite{watanabe2020chime}, which leveraged close-talk microphones and external datasets for RIRs and noises, namely SLR26~\cite{ko2017study} and MUSAN~\cite{snyder2015musan}. We also used GSS-enhanced data obtained from the whole CHiME-6 training set.

\vspace{-9pt}
\subsection{Experimental setup}
\vspace{-0.1cm}
All of our systems are implemented with the ESPnet~\cite{watanabe2018espnet} toolkit. We follow the setup of the baseline ASR system to build our systems, which consist of a WavLM frontend, a 12-layer Transformer encoder, and a 6-layer Transformer decoder. The dimensions of MHSA and FFN layers are set to 256 and 2048, respectively.
All of our systems are trained on the full CHiME-7 train sets with oracle segmentation and processed following the procedures outlined in Section~\ref{sec:data_process}, 276 hours in total. Besides, all of our systems are rescored by a Transformer-based language model (LM) trained on a combination of the CHiME-7 and LibriSpeech~\cite{panayotov2015librispeech} corpora.
During the training, we freeze the parameters of the WavLM. 
AFE is initialized with a well-trained ASR system that utilizes a 4-layer bi-directional GRU encoder trained solely with the CTC loss. 
The spatial feature cosIPD is extracted with window length, frameshift, and STFT length are 32ms, 16ms, and 512, respectively. 
\vspace{-9pt}
\subsection{Results and discussion}
\vspace{-0.1cm}
Table~\ref{tab:all_results} presents the DA-WER results of our proposed methods on the CHiME-7 Dev and Eval sets. As shown in Table~\ref{tab:all_results}, all of our proposed methods outperform than baseline. 
Among all proposed methods, FGCS yields the most significant benefits. This is because FGCS selects the most correlated frame-level features with the GSS audio feature from the multi-channel audio features. 
Furthermore, the combined utilization of CGCS and GRC contributes significantly more to the reduction in DA-WER compared to using CGCS alone, making it the second most effective approach after FGCS. This phenomenon can be attributed to the direct summation of the original features with the features selected through CGCS outside the encoder. This process may reintroduce numerous features that are unfavorable for ASR. The incorporation of GRC effectively addresses this issue by allowing for a selective reintroduction of the original features.
% The incorporation of cosIPD features noticeably enhances the MFCCA's capacity to perceive spatial information.
The incorporation of cosIPD features into MFCCA leads to a noticeable reduction in DA-WER. This improvement can be attributed to the direct modeling of spatial features by MFCCA, which significantly enhances its ability to perceive spatial information.
Replacing the convolution fusion module in MFCCA with the U-Net-based convolution fusion module allows the ASR system to compress the channel dimension while retaining a greater amount of channel-specific information.
By incorporating all the proposed methods into the ASR system, we achieve the best performance. The DA-WER shows a relative reduction of 24.6\% on the Dev sets and 34.1\% on the Eval sets.
Finally, we utilize the ROVER~\cite{fiscus1997post} to combine the results of all the aforementioned ASR systems. This leads to a relative reduction of 30.2\% in DA-WER on the Dev sets and 40.1\% on the Eval sets.
\begin{figure}[t]
  \centering
  \includegraphics[width=1.0\linewidth]{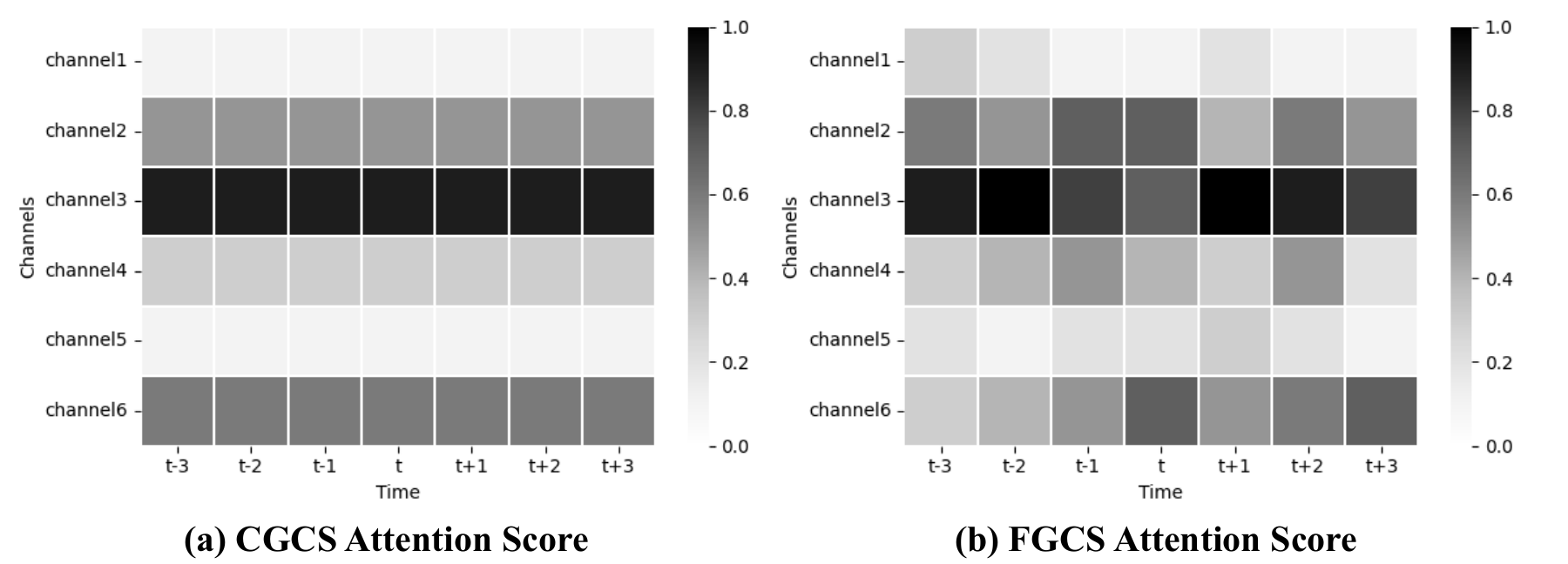}
  \caption{A visualization of attention scores for CGCS and FGCS.}
  \vspace{-18pt}
   \label{fig:attn_score}
\end{figure}
\vspace{-9pt}
\subsection{Visualization of CGCS and FGCS}
\vspace{-0.1cm}
To analyze the behavior of CGCS and FGCS, Fig.~\ref{fig:attn_score} respectively visualizes the attention scores of CGCS and FGCS. 
CGCS can assign different scores based on the richness of semantic information across different channels.
Channels with richer semantic information are assigned higher scores. 
FGCS assigns different scores to each frame of the multi-channel audio features based on the frame-level similarity between the multi-channel audio features and the GSS audio features. 
Frames in the multi-channel audio features that exhibit higher similarity with the GSS audio features are assigned higher scores.
Channels that contain richer semantic information tend to have more frames with high similarity to the GSS audio features.
As shown in Fig.~\ref{fig:attn_score}, the third channel has the most abundant semantic information and also has the highest quantity of frames that are most similar to the GSS audio features.
% the 1-st and 5-th channel have the least semantic information and also contain a lower quantity of frames that exhibit low similarity to the GSS audio features.
\begin{table}[]
    \caption{The Macro DA-WER results of other ASR systems on the Eval sets. ISS: Improved Speech Separation; FT: WavLM Finetuning; DA: Data Augmentation; ID: Improved Decoding; IAM: Improved Acoustic Model.}
    \label{tab:compare_results}
    \centering
\scalebox{1}{
    \begin{tabular}{lccllcc}
    \toprule
    \textbf{ASR System}  & \textbf{ISS} & \textbf{FT} & \textbf{DA} & \textbf{ID} & \textbf{IAM} & \textbf{Eval (\%)} \\ \midrule
    1$^{st}$ Rank~\cite{wang2023ustc} & \ding{52}      & \ding{52}  & \ding{52}  & \ding{56}  & \ding{56}   & 16.8 \\
    2$^{nd}$ Rank~\cite{Prisyach2023stcon} & \ding{52}      & \ding{52}  & \ding{56}  & \ding{52}  & \ding{56}   & 18.3 \\ \midrule
    Ours        & \ding{56}      & \ding{56}  & \ding{56}  & \ding{56}  & \ding{52}   & 20.0 \\ \bottomrule
    \vspace{-27pt}
    \end{tabular}
}
\end{table}

% \begin{table}[]
%     \caption{The Macro DA-WER results of other ASR systems on the Eval sets.}
%     \label{tab:compare_results}
%     \centering
% \begin{tabular}{lcccc}
% \toprule
% ASR System  & Data (hrs) & Fusion & FT & Eval Macro (\%) \\ \midrule
% 1$^{st}$ Rank~\cite{wang2023ustc} & 4500      & \ding{52}            & \ding{52}  & 16.8       \\ 
% 2$^{nd}$ Rank~\cite{Prisyach2023stcon} & -         & \ding{52}            & \ding{52}  & 18.3       \\ \midrule
% Ours        & 276       & \ding{52}            & \ding{56}  & 20.0       \\ \bottomrule
% \vspace{-27pt}
% \end{tabular}
% \end{table}

\vspace{-9pt}
\subsection{Compared with other systems}
\vspace{-0.1cm}
% Table~\ref{tab:compare_results} shows the comparison between the top two systems and ours. CACGMM, FT, DA, ID, and IAM respectively represent the CACGMM data processing method~\cite{ito2016complex}, WavLM Finetuning, Data Augmentation, Improved Decoding, and Improved Acoustic Model.
As shown in Table~\ref{tab:compare_results}, our system currently ranks third in the CHiME-7 acoustic robustness sub-track, but there is still a gap compared to the top two systems. 
% They both use the CACGMM data processing method and finetune WavLM.
To tackle the challenge of high overlapped speech ratios in the CHiME-7 dataset, they both employ additional data processing algorithms for speech separation, such as CACGMM~\cite{ito2016complex}, alongside GSS. 
Moreover, they both finetune WavLM to harness its capabilities further.
The training data scale for the first-ranked system is 4,500 hours after data augmentation. The second-ranked system uses joint CTC/Attention decoding and Kaldi decoding with TLG graph and rescores ASR results using 3-gram LM, Transformer-LM, and AWD-LSTM-LM~\cite{merity2017regularizing} trained on a combination of the CHiME-7, LibriSpeech, and WSJ corpora.
However, neither of them improves the acoustic model, using only the original Conformer, E-Branchformer, and Zipformer.
In contrast, our primary focus lies in enhancing the performance of the acoustic model and outcomes reflect the effectiveness of our proposed methods.
In the future, we will enhance our ASR system to address the issue of overlapped speech. Moreover, we will attempt to unfreeze WavLM and conduct joint training with our ASR system. Finally, we will explore more effective data preprocessing, data augmentation, and decoding methods to achieve lower DA-WER.
% All the results are obtained after model fusion. Compared to the first-place result, our training data is significantly smaller. We train our system using only the CHiME-7 train sets, while the first-place system extended the CHiME-7 train sets using different data processing methods and also make use of Librispeech and VoxCeleb1\&2. 
% Furthermore, while all three systems use WavLM as a feature extractor, the top two systems unfreeze WavLM and jointly train it with ASR, whereas we freeze WavLM all the time.
% Finally, the data processing methods used by the top two systems are more complex and effective than ours.
\vspace{-9pt}
\section{Conclusions}
\vspace{-0.1cm}
% In this work, we propose a multi-channel ASR system that excels consistently across diverse array topologies. Two attention-based automatic channel selection modules with GRC select subsets from the multi-channel signals that are favorable for ASR. The addition of spatial features to MFCCA improves its spatial awareness. The U-Net-based convolution fusion module not only reduces the channel dimension but also preserves channel information better. 
% Demonstrated the effectiveness of all proposed methods through experimental validation. 
In this work, we propose an ASR system that excels consistently across various array topologies each with multiple recording devices. The ASR system includes two attention-based automatic channel selection modules with GRC, improved MFCCA by integrating spatial features, and a multi-layer convolution fusion module inspired by the U-Net architecture.
Demonstrated the effectiveness of all proposed methods through experimental validation.
% Our ASR system achieves a relative reduction of 24.6\% and 34.1\% in DA-WER on the Dev and Eval sets compared to the CHiME-7 baseline, respectively. 
% After combining the results of multiple ASR systems using ROVER, 
Our ASR system achieves a relative reduction in DA-WER on the Dev and Eval sets is 30.2\% and 40.1\%, respectively. 
\newpage
% References should be produced using the bibtex program from suitable
% BiBTeX files (here: strings, refs, manuals). The IEEEbib.bst bibliography
% style file from IEEE produces unsorted bibliography list.
% -------------------------------------------------------------------------
% \scriptsize
% \vspace{-0.2cm}
\bibliographystyle{IEEEbib}
\bibliography{strings,refs}

\end{document}